\documentclass[11pt]{revtex4}
\usepackage{color}
\usepackage{amsmath,amssymb}
\usepackage[dvipdfm]{graphicx}
\usepackage{cancel}

\allowdisplaybreaks

\begin{document}
\title{Interaction solutions for mKP equation with nonlocal symmetry reductions and CTE method}

\author{Bo Ren$^{1}$\footnote{E-mail: renbosemail@gmail.com.}}

\affiliation{$^1$Institute of Nonlinear Science, Shaoxing University, Shaoxing 312000, China }

\date{\today $\vphantom{\bigg|_{\bigg|}^|}$}

\begin{abstract}

The nonlocal symmetries for the modified Kadomtsev-Petviashvili (mKP) equation are obtained with the truncated Painlev\'{e} method. The nonlocal symmetries can be localized to the Lie point symmetries by introducing auxiliary dependent variables. The finite symmetry
transformations and similarity reductions related with the nonlocal symmetries are computed.
The multi-solitary wave solution and interaction solutions among a soliton and cnoidal waves of the mKP equation are presented.
In the meanwhile, the consistent tanh expansion (CTE) method is applied to the mKP equation. The explicit interaction solutions among
a soliton and other types of nonlinear waves such as cnoidal periodic waves and multiple
resonant soliton solutions are given.

\end{abstract}

\maketitle



\section{Introduction}

The study of nonlinear integrable systems is one of the most important subjects in modern physics and
nonlinear science. The explicit solutions of the nonlinear integrable equations are very important due to
their wide applications in explaining physical phenomena.
The most effective methods are the inverse scattering transformation \cite{ism},
bilinear form \cite{bili}, symmetry reduction \cite{Olver}, Darboux transformation \cite{darx}, Painlev\'{e} analysis method \cite{pain}, B\"{a}klund transformation (BT) \cite{bt}, separated variable method \cite{vds}, etc. For these methods, the interaction solutions between solitons and nonlinear waves are difficult to obtain. Recently, Lou and his colleagues proposed the
localization procedure related with the nonlocal symmetry \cite{darb,gaox} and the consistent tanh expansion (CTE) method \cite{gaox,cxp} to find interaction solutions. Some interesting interaction solutions between a soliton
and the cnoidal waves, Painlev\'{e} waves, Airy waves, Bessel waves are generated with the methods \cite{darb,gaox,cxp,cheny,sbt,cte,Interaction,wang,scho}.

In this letter, we focus on the modified Kadomtsev-Petiashvili (mKP) equation \cite{Konopelchenko}
\begin{subequations}\label{mkp}
\begin{eqnarray}
&u_{t} - \frac{1}{4} u_{xxx} - \frac{3}{4} w_{y} + \frac{3}{2}u^2 u_{x} + \frac{3}{2} u_{x} w = 0,\\
& \hspace{0.8cm} u_y = w_x,
\end{eqnarray}
\end{subequations}
which describes water waves in $(x, y)$ plane when the nonlinearity
is higher than for the KP equation.
The integrable properties of this equation such as the Lax pair \cite{Konopelchenko,cheng}, Darboux transformation \cite{invers,deng} and explicit solutions \cite{geng,dai} have been obtained. From the mKP equation, a new integrable system is given by means of an asymptotically exact reduction
method \cite{ren}.

The paper is organized as follows. In section 2, the nonlocal
symmetries for the mKP equation are obtained with the truncated Painlev\'{e} method. To solve the initial value problem of the nonlocal symmetries,
the nonlocal symmetries are localized by prolongation the mKP equation. The finite symmetry transformations are obtained by solving the initial value problem of the Lie's first principle. The multi-solitary wave solution of the mKP equation is obtained using the finite symmetry transformations.
In section 3, the group invariant solutions related to the nonlocal symmetries are studied with the symmetry reduction method. The corresponding explicit interaction solutions among a soliton and cnoidal waves are given.
Section 4 is devoted to the CTE approach for the mKP equation.
Some exact interaction solutions among different nonlinear excitations such as
cnoidal periodic waves and multiple resonant soliton solutions are explicitly given.
The last section is a simple summary and discussion.

\section{Nonlocal symmetries of the mKP equation and explicit solutions}

To construct the BT of the mKP equation, we truncate the Laurent series as \cite{pain}
\begin{align}\label{bt}
u=\frac{u_0}{\phi}+ {u_1}, \hspace{1cm} w=\frac{w_0}{\phi}+{w_1},
\end{align}
where the function $\phi(x, y, t)=0$ is the equation of singularity manifold, the
functions $u_0$, $u_1$, $w_0$ and $w_1$ are determined by substituting of expansion \eqref{bt} into \eqref{mkp} and balancing all coefficients of each power of $\phi$ independently. Then, we get
\begin{align}\label{uw0}
u_0 = \phi_x, \hspace{1.6cm}  w_0=\phi_{y}, \hspace{0.2cm}
\end{align}
\begin{align}\label{uw1}
u_1 = -\frac{1}{2} \Bigl( \frac{\phi_{xx}}{\phi_x} + \frac{\phi_y}{\phi_x}\Bigr), \hspace{0.6cm} w_1= - \frac{2}{3}\frac{\phi_t}{\phi_x} - \frac{\phi_{xy}}{2\phi_x} + \frac{\phi_{xxx}}{6\phi_x} + \frac{1}{4} \Bigl(\frac{\phi_{y}^2}{\phi_x^2}- \frac{\phi_{xx}^2}{\phi_x^2}\Bigr)  .
\end{align}
Substituting the expressions \eqref{bt}, \eqref{uw0} and \eqref{uw1} into \eqref{mkp}, the field $\phi$ satisfy the following Schwarzian mKP form
\begin{align}\label{sch}
4\Bigl(\frac{\phi_t}{\phi_x}\Bigr)_x - \frac{\partial}{\partial x} \{\phi; x\} - 3 \Bigl(\frac{\phi_y}{\phi_x}\Bigr)_y -
\frac{3}{2} \Bigl(\frac{\phi_y^2}{\phi_x^2}\Bigr)_x =0 ,
\end{align}
where $\{\phi;x\}=\frac{\partial}{\partial x}\bigl(\frac{\phi_{xx}}{\phi_x}\bigr)-\frac{1}{2}\bigl(\frac{\phi_{xx}}{\phi_x}\bigr)^2$ is the
Schwarzian derivative.
For the BT \eqref{bt}, two pairs of functions $u, w$ and $u_1, w_1$ satisfy \eqref{mkp}. The latter solutions $u_1, w_1$ are related to $\phi$  with \eqref{uw1}.

The nonlocal symmetries of the mKP equation \eqref{mkp} can read out from the BT~\cite{gaox}
\begin{align}\label{rsnon}
\sigma^u=\phi_{x}, \hspace{1.2cm}\sigma^w=\phi_{y}.
\end{align}
The nonlocal symmetries \eqref{rsnon} are the residual of the singularity manifold $\phi$. Thus, this nonlocal symmetries are called as the residual
symmetries (RS)~\cite{gaox}. Besides, this nonlocal symmetries can also be obtained from the Schwarzian form \eqref{sch} \cite{schwr}. The Schwarzian form \eqref{sch} is invariant under the M\"{o}bious transformation
\begin{align}
\phi \rightarrow \frac{a\phi+b}{c\phi+d},\hspace{1cm} ac\neq bd,
\end{align}
which means \eqref{sch} possesses the symmetry $\sigma^\phi= - \phi^2$ in special case $a=d=1,\, b=0,\, c=\epsilon.$
The nonlocal symmetries \eqref{rsnon} will be obtained with substituting the M\"{o}bious transformation symmetry $\sigma^\phi$
into the linearized equation of \eqref{uw1}.

For the nonlocal symmetries \eqref{rsnon}, the corresponding initial value problem is
\begin{align}\label{inp}
&\frac{d\overline{u}}{d\epsilon}=\phi_{x}, \hspace{1.6cm} \overline{u}|_{\epsilon=0}=u,\\ \nonumber
&\frac{d\overline{w}}{d\epsilon}=\phi_{y}, \hspace{1.6cm} \overline{w}|_{\epsilon=0}=w.
\end{align}
It is difficult to solve the initial value problem of the Lie's first principle \eqref{inp} due to the intrusion of the function $\phi$ and
its differentiations \cite{gaox}.
To solve the initial value problem \eqref{inp}, we prolong the mKP system \eqref{mkp} such that RS become the local Lie point
symmetries for the prolonged system.
By localization the nonlocal symmetries \eqref{rsnon}, the potential fields of $\phi$ are introduced as
\begin{align}\label{poent}
\phi_x = g,
\end{align}
\begin{align}\label{poen}
\phi_y = h.
\end{align}
It is easy to verify that the local Lie point symmetries for the prolonged systems \eqref{mkp}, \eqref{uw1}, \eqref{poent} and \eqref{poen} read as
\begin{align}\label{nores}
\sigma^u = g, \hspace{1cm}
\sigma^w = h, \hspace{1cm}
\sigma^\phi=-\phi^2, \hspace{1cm}
\sigma^g=-2\phi g,\hspace{1cm}
\sigma^h=-2\phi h.
\end{align}
Correspondingly, the initial value problem becomes
\begin{align}\label{ini}
& \frac{d\overline{u}}{d\epsilon} = g, \hspace{2.7cm} \overline{u}|_{\epsilon=0}=u,\nonumber \\
& \frac{d\overline{w}}{d\epsilon} = h, \hspace{2.6cm} \overline{w}|_{\epsilon=0}=w,\nonumber \\
& \frac{d\overline{\phi}}{d\epsilon} = -\phi^2, \hspace{2.15cm} \overline{\phi}|_{\epsilon=0}=\phi,\\
& \frac{d\overline{g}}{d\epsilon} = -2\phi g , \hspace{2.05cm} \overline{g}|_{\epsilon=0}=g,\nonumber\\
& \frac{d\overline{h}}{d\epsilon} = -2\phi h, \hspace{2.0cm} \overline{h}|_{\epsilon=0}=h.\nonumber
\end{align}
The solution of the initial value problem \eqref{ini} for the enlarged mKP system \eqref{mkp}, \eqref{uw1}, \eqref{poent} and \eqref{poen} can be written as
\begin{align}\label{fsou}
\overline{u} = u + \frac{\epsilon g}{\epsilon \phi+1}, \hspace{0.41cm}
\overline{w} = w + \frac{\epsilon h}{\epsilon \phi+1}, \hspace{0.41cm}
\overline{\phi}= \frac{\phi}{\epsilon \phi+1}, \hspace{0.41cm}
\overline{g}=\frac{g}{(\epsilon \phi+1)^2},\hspace{0.41cm}
\overline{h}=\frac{h}{(\epsilon \phi+1)^2}.
\end{align}
Using the finite symmetry transformations \eqref{fsou}, one can obtain a new solution from any
initial solution. For example, we take the trivial solution $u = w = 0$ for \eqref{mkp}.
The multi-solitary wave solution for \eqref{uw1} is supposed as \cite{fusi}
\begin{align}\label{nsol}
\phi = 1 + \sum^{N}_{n=1} \exp(k_n x + l_n y + \omega_n t),
\end{align}
where $k_n, l_n$ and $\omega_n$ are arbitrary constants.
The multi-solitary wave solution \eqref{nsol} is the solution of \eqref{uw1} and \eqref{sch} only with the relations
\begin{eqnarray}
l_n=-k_n^2, \hspace{1.2cm}\omega_n = k_n^3.
\end{eqnarray}
A solution of equation \eqref{mkp} presents in the following form using \eqref{poent}, \eqref{poen} and \eqref{fsou}
\begin{subequations}\label{nsolu}
\begin{eqnarray}
& u = \frac{ \sum^{N}_{n=1} \epsilon k_n\exp(k_n x + l_n y + \omega_n t)}{-1 + \epsilon +  \sum^{N}_{n=1}\epsilon \exp(k_n x + l_n y + \omega_n t)}, \,\,\\
& w = -\frac{ \sum^{N}_{n=1} \epsilon k_n^2\exp(k_n x + l_n y + \omega_n t)}{-1 + \epsilon + \sum^{N}_{n=1} \epsilon \exp(k_n x + l_n y + \omega_n t)}.
\end{eqnarray}
\end{subequations}
Figure 1 shows the three-solitary wave solution of the fields $u$ and $w$ with the parameters $n=3,\, \epsilon=-\frac{1}{4},\, k_1=-1,\, k_2=-2,\, k_3=3$.
\input epsf
\begin{figure}
\centering
\rotatebox{270}{\includegraphics[width=4.75cm]{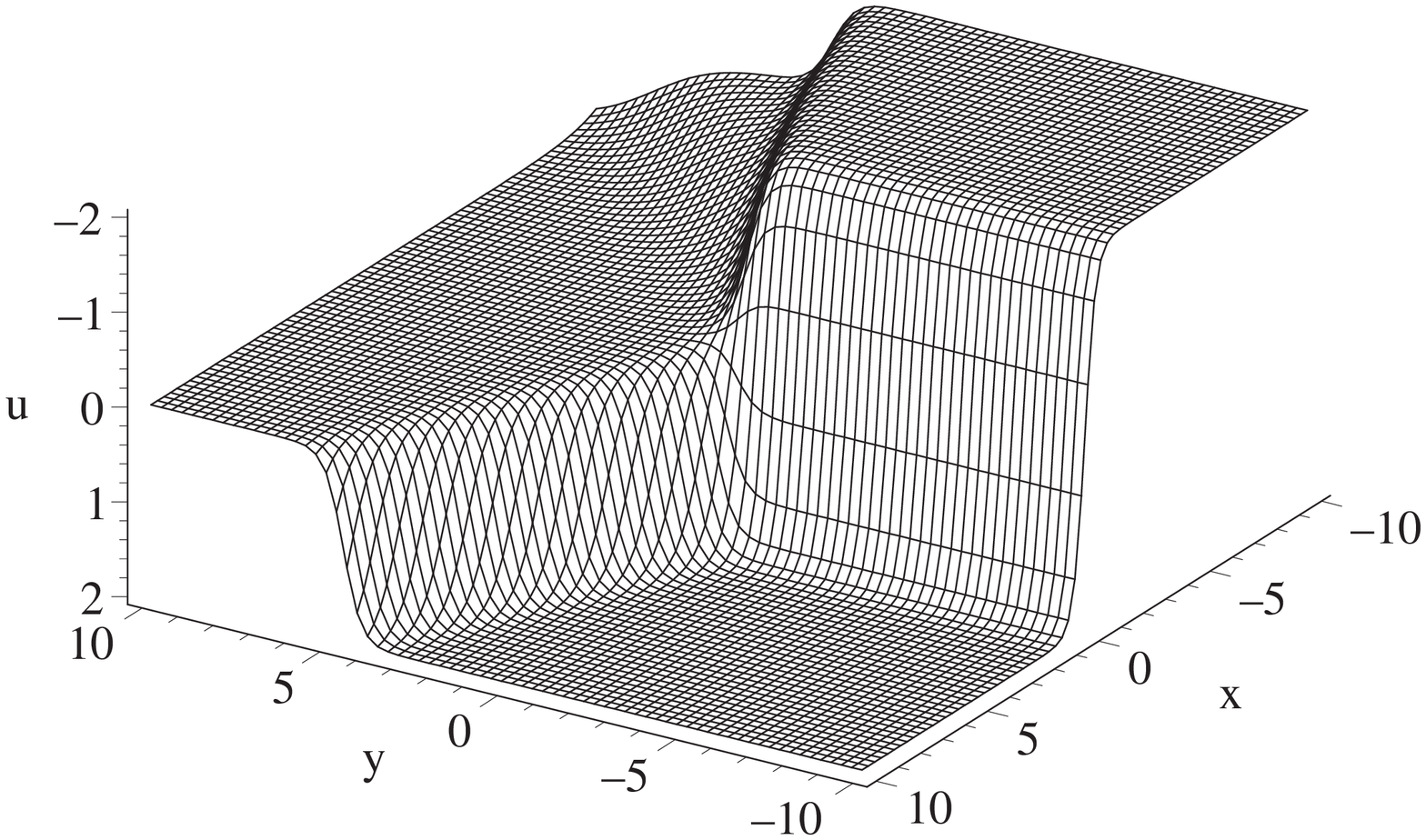}}
\rotatebox{270}{\includegraphics[width=4.95cm]{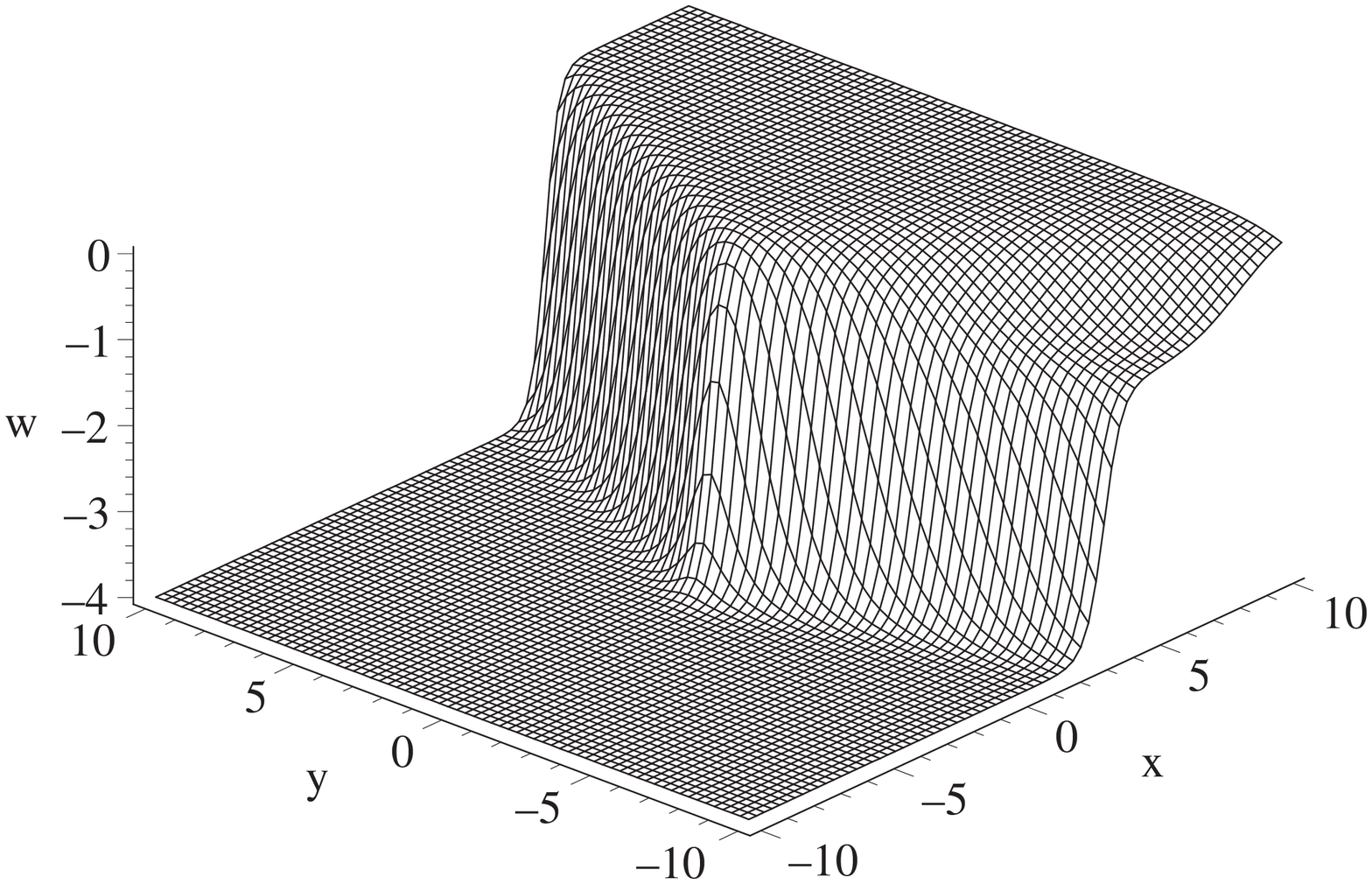}}
\caption{Three-solitary wave solution for the fields $u$ and $w$ with the parameters $n=3,\, \epsilon=-\frac{1}{4},\, k_1=-1,\, k_2=-2,\, k_3=3$ and $t=0$ respectively.}
\end{figure}

\section{Similarity reductions with the nonlocal symmetries}

In this section, the symmetry reductions related to the nonlocal symmetries will be discussed.
According to the standard Lie symmetry approach, the Lie point symmetries for the prolonged systems possess the form
\begin{align}\label{symmf}
&\sigma^u = X u_x + Y u_y + T u_t - U, \nonumber \\
&\sigma^w = X w_x + Y w_y + T w_t - W, \nonumber \\
&\sigma^\phi = X \phi_x + Y \phi_y + T \phi_t - \Phi, \\
&\sigma^g = X g_x + Y g_y + T g_t - G, \nonumber \\
&\sigma^h = X h_x + Y h_y + T h_t - H, \nonumber
\end{align}
where $X, Y, T, U, W, \Phi, G, H$ are functions of $x, t, u, w, \phi, g, h$.

The symmetries $\sigma^k \,(k=u, w, \phi, g, h)$ are defined as the solution of the linearized equations of the prolonged systems \eqref{mkp} \eqref{uw1}, \eqref{poent} and \eqref{poen}
\begin{subequations}\label{fosym}
\begin{eqnarray}
& \sigma^u_{\,t} - \frac{1}{4} \sigma^u_{\,xxx} - \frac{3}{4}\sigma^w_{\,y} + 3\sigma^u u u_x + \frac{3}{2} u^2 \sigma^u_{\,x} + \frac{3}{2}\sigma^u_{\,x}w +\frac{3}{2}\sigma^w u_x = 0, \\
& \sigma^u_{\,y} - \sigma^w_{\,x} = 0,\\
& \sigma^{u} + \frac{1}{2} \frac{\sigma^\phi_{\,y}}{\phi_x} + \frac{1}{2} \frac{\sigma^\phi_{\,xx}}{\phi_x} - \frac{1}{2} \frac{\sigma^\phi_{\,x}\phi_{y}}{\phi_x^2} - \frac{1}{2} \frac{\sigma^\phi_{\,x}\phi_{xx}}{\phi_x^2}=0,\\
& \sigma^{w} + \frac{4\sigma^\phi_{\,t} + 3\sigma^{\phi}_{\,xy} - \sigma^{\phi}_{\,xxx}}{6\phi_x}
- \frac{2}{3} \frac{\sigma^\phi_{\,x}\phi_{t}}{\phi_x^2} - \frac{\sigma^\phi_{\,y}\phi_{y}}{2\phi_x^2} - \frac{\sigma^\phi_{\,x}\phi_{xy}}{2\phi_x^2} + \frac{\sigma^\phi_{\,x}\phi_{xxx}}{6\phi_x^2} + \frac{\sigma^\phi_{\,xx}\phi_{xx}}{2\phi_x^2} + \frac{\sigma^\phi_{\,x}\phi_y^2 - \sigma^\phi_{\,x}\phi_{xx}^2}{2\phi_x^3} =0,\\
& \sigma^\phi_{\,x} - \sigma^g = 0,\\
& \sigma^\phi_{\,y} - \sigma^h = 0.
\end{eqnarray}
\end{subequations}
That is to say, the prolong systems are invariant under transformations
\begin{align}
u\rightarrow u + \epsilon \sigma^u, \hspace{0.6cm} w\rightarrow w + \epsilon \sigma^w, \hspace{0.6cm}
\phi\rightarrow \phi + \epsilon \sigma^\phi, \hspace{0.6cm} g \rightarrow g + \epsilon \sigma^g, \hspace{0.6cm}
h\rightarrow h + \epsilon \sigma^h, \hspace{0.6cm}
\end{align}
with the infinitesimal parameter $\epsilon$.

Substituting \eqref{symmf} into the symmetry equations \eqref{fosym} and requiring $u, w, \phi, g, h$ to satisfy the prolonged systems, the determining equations are obtained with collecting the coefficients of $u, w, \phi, g, h$ and its derivatives.
The infinitesimals $X$, $Y$, $T$, $U$, $W$, $\Phi$, $G$ and $H$ are given by solving the determining equations
\begin{align}\label{dete}
& X= \frac{1}{3}f_{1t} x + \frac{2}{3} f_{2t}y +\frac{2}{9}f_{1tt}y^2+f_3, \hspace{0.2cm}Y= \frac{2}{3}f_{1t}y+f_2, \hspace{0.2cm} T = f_1 t + C_1, \hspace{0.2cm}  G = 2C_2\phi g + C_3g -\frac{1}{3}f_{1t} g,   \nonumber \\
&U = - \frac{1}{3}f_{1t} u - C_2 g + \frac{2}{9}f_{1tt} y + \frac{1}{3} f_{2t},
\hspace{1cm} H=2C_2\phi h + C_3 h -\frac{2}{3} f_{1t}h - \frac{2}{3}f_{2t}g -\frac{4}{9}f_{1tt}gy, \\
&W= -\frac{2}{3}f_{1t}w - \frac{2}{3}f_{2t}u - \frac{4}{9}f_{1tt}yu -C_2h +\frac{2}{9}f_{1tt}x +\frac{4}{27}f_{1ttt}y^2 + \frac{4}{9}f_{2tt}y+\frac{2}{3}f_{3t}, \hspace{0.1cm} \Phi = C_2\phi^2 + C_3 \phi + C_4, \nonumber
\end{align}
where $f_1, f_2$ and $f_3$ are arbitrary functions of $t$ and $C_1, C_2, C_3$ and $C_4$ are arbitrary constants. It is well known that whence a symmetry is known, the related group invariant solutions can be naturally obtained with the symmetry constraint condition $\sigma^k=0$ defined by \eqref{symmf}. It is
equivalent to solving the following characteristic equations
\begin{align}\label{chara}
\frac{dx}{X} = \frac{dy}{Y} = \frac{dt}{T}=\frac{du}{U}=\frac{dw}{W}=\frac{d\phi}{\Phi}=\frac{dg}{G}=\frac{dh}{H}.
\end{align}
To solve the characteristic equations, two special cases are listed in the following.

{\bf Case I.} $f_1=0$.
In this case, without loss of generality, we rewrite the arbitrary
functions $f_2$ and $f_3$ as
\begin{align}
f_2=C_1 m_{1t}, \hspace{1cm} f_3=C_1 m_{2t},
\end{align}
where $m_1$ and $m_2$ being arbitrary functions of $t$.
We can find the similarity solutions after solving out the characteristic equations \eqref{chara}
\begin{subequations}\label{slosc}
\begin{align}
& \phi =  \frac{\Delta}{2C_2} M - \frac{C_3}{2C_2}, \hspace{1cm}\Delta=\sqrt{4C_2C_4-C_3^2},\\
& g = G(\sec M)^2,\\
& h = (H  - \frac{2}{3}m_{1t} G)(\sec M)^2,\\
& u = U + \frac{2C_2}{\Delta} G (\arctan M -M) - \frac{C_2}{C_1} t G + \frac{1}{3}m_{1t},\\
& w = W + \frac{2C_2}{\Delta} H (\arctan M -M) - \frac{4C_2}{3\Delta} m_{1t} G (\arctan M -M) - \frac{C_2}{C_1}t H \\
& \hspace{0.65cm}  - \frac{2}{3}m_{1t}U + \frac{2C_2}{3C_1}tm_{1t}G + \frac{4}{9}m_{1tt}y - \frac{1}{9}m_{1t}^2 + \frac{2}{3}m_{2t}, \nonumber
\end{align}
\end{subequations}
with the similarity variables $\xi=x-\frac{2}{3}m_{1t}y - m_2$ and $\eta=y-m_1$ and $M=\tan \bigl(\frac{\Delta}{2C_1} (\Phi+t)\bigr)$ for similarity. Substituting \eqref{slosc} into \eqref{poent}, \eqref{poen}, \eqref{uw1} and \eqref{sch}, the invariant functions $G$, $H$, $U$, $W$ and $\Phi$ satisfy the reduction systems
\begin{subequations}\label{slfc}
\begin{align}
& G = \frac{\Delta^2}{4C_1C_2} \Phi_\xi,\\
& H = \frac{\Delta^2}{4C_1C_2} \Phi_\eta,\\
& U = \frac{\Delta^2}{4C_1^2} t\Phi_\xi - \frac{\Delta}{2C_1}\Phi_\xi \arctan M - \frac{\Phi_\eta + \Phi_{\xi\xi} }{2\Phi_\xi},\\
& W = \frac{2}{9}m_{1t}^{\,2} + \frac{\Delta^2}{12C_1^2}\Phi_\xi^{\,2} + \frac{\Delta^2}{4C_1^2} t\Phi_\eta +
\frac{\Phi_{\xi\xi\xi} - 3\Phi_{\xi\eta} -4}{6\Phi_\xi} + \frac{\Phi_\eta^{\,2}-\Phi_{\xi\xi}^{\,2}}{4\Phi_\xi^{\,2}} -
\frac{\Delta}{2C_1}\Phi_\eta \arctan M,\\
&\frac{\Delta^2}{C_1^2} \Phi_{\xi\xi} \Phi_{\xi}^4 + \Phi_\xi^2\Phi_{\xi\xi\xi\xi} + 3 \Phi_{\eta\eta}\Phi_{\xi}^2 - 4\Phi_{\xi} \Phi_{\xi\xi}\Phi_{\xi\xi\xi}
+3 \Phi_{\xi\xi}^3 - 3\Phi_{\xi\xi}\Phi_{\eta}^2 + 4 \Phi_{\xi}\Phi_{\xi\xi} =0.
\end{align}
\end{subequations}
It is obvious that once the solutions $\Phi$ are solved out with (24e), the fields for $G$, $H$, $U$ and $W$ can be solved out directly from (24a)-(24d). The explicit solutions for mKP \eqref{mkp} are
immediately obtained by substituting $\Phi$, $G$, $H$, $U$ and $W$ and into \eqref{slosc}.

For instance, it has a trivial solution $\Phi=\xi+\eta$ for the system (24e). The exact solution for mKP \eqref{mkp} express as
\begin{align}\label{solu1}
& u= -\frac{\Delta}{2C_1}M + \frac{1}{3}m_{1t} -\frac{1}{2},\\
& w= \Bigl(-\frac{1}{3}m_{1t} - \frac{1}{2}\Bigr) \frac{\Delta}{C_1}M + \frac{4}{9} m_{1tt} (m_1+\eta) + 9m_{1t}^2 +\frac{1}{3}m_{1t} +\frac{2}{3}m_{2t} + \frac{\Delta^2}{12C_1^2} - \frac{5}{12}, \nonumber
\end{align}
which is a nontrivial solution of the mKP equation.
Besides, for the reduction equation (24e), its cnodial wave solution is given
\begin{align}\label{simp}
\Phi_{1X}^2 - \frac{\Delta^2}{C_1^2} \Phi_1^4 + 2B \Phi_1^3 -2A \Phi_1^2 + 4 \Phi_1 =0, \hspace{0.45cm} \Phi_1=\Phi_X, \hspace{0.45cm} \Phi(\xi,\eta)= \Phi(\xi+a\eta)= \Phi(X),
\end{align}
where arbitrary constants $a, A, B$. The solution for \eqref{simp} can be explicitly expressed by Jacobi elliptic functions
\begin{align}
\Phi_{1} = - \frac{r_1r_3 S^2}{r_1 S^2 -r_1+r_3}, \hspace{0.6cm} S=\mathrm{sn}\left(\frac{\Delta}{2C_1}\sqrt{r_2(r_1-r_3)}X, m \right), \hspace{0.6cm}m=\frac{r_1(r_2-r_3)}{r_2(r_1-r_3)},
\end{align}
where $S$ is the Jacobi elliptic function with the modulus $m$ and the arbitrary constants $C_1, C_2$ and $\frac{\Delta}{C_1}$ have been re-expressed by
\begin{align}
A=\frac{2(r_1r_2+r_1r_3+r_2r_3)}{r_1r_2r_3}, \hspace{0.7cm} B= \frac{2(r_1+r_2+r_3)}{r_1r_2r_3},\hspace{0.7cm} \frac{\Delta}{C_1}=\frac{2}{\sqrt{r_1r_2r_3}}. \nonumber
\end{align}
Correspondingly, the field $\Phi$ has the form
\begin{align}\label{jorb}
\Phi = \frac{2r_3C_1}{\Delta\sqrt{r_2(r_1-r_3)}} \Bigl(E_{\pi} (S, \frac{r_1}{r_1-r_3}, m) - E_F(S, m)\Bigr),
\end{align}
where $E_F$ and $E_{\pi}$ are the first and third incomplete elliptic integrals. It is clear that the exact solution
for the mKP equation denotes the interaction between a soliton and cnoidal periodic waves.

{\bf Case II.} $f_1=f_2=C_1=0$.
We find the similarity solutions after solving out the characteristic equations \eqref{chara}
\begin{subequations}\label{scon}
\begin{align}
& \phi =  \frac{\Delta}{2C_2} \tan\Bigl( \frac{\Delta }{2f_3} (\phi{'}+x)\Bigr) - \frac{C_3}{2C_2},\\
& g = g' \sec\Bigl( \frac{\Delta }{2f_3} (\phi{'}+x)\Bigr)^2,\\
& h = h' \sec\Bigl( \frac{\Delta }{2f_3} (\phi{'}+x)\Bigr)^2,\\
& u = u{'} - \frac{2C_2}{\Delta} g' \tan\Bigl( \frac{\Delta }{2f_3} (\phi{'}+x)\Bigr),\\
& w = w' + \frac{2}{3} \frac{f_3'}{f_3} (\phi{'}+x) - \frac{2C_2}{\Delta} h' \tan\Bigl( \frac{\Delta }{2f_3} (\phi{'}+x)\Bigr),
\end{align}
\end{subequations}
where the group invariant functions $\phi{'}=\phi{'}(y, t)$, $g{'}=g{'}(y, t)$, $h{'}=h{'}(y, t)$, $u{'}=u{'}(y, t)$ and $w{'}=w{'}(y, t)$. Substituting \eqref{scon} into \eqref{uw1}, \eqref{poent}, \eqref{poen}, the invariant functions $\phi{'}, g{'}, h{'}, u{'}$ and $w{'}$ satisfy the reduction systems
\begin{subequations}\label{slff}
\begin{align}
& \phi' = -\frac{2f_{3t}}{3f_3} y^2 - 2f_4y +f_5\\
& g' = \frac{\Delta^2}{4C_2f_3},\\
& h' = \frac{\Delta^2\phi{'}_y}{4C_2f_3},\\
& u' = \frac{2f_{3t}}{3f_3} y + f_4,\\
& w' = \frac{4f_{3tt}}{9f_3} y^2 + \frac{4(f_3f_4)_t}{3f_3} y  + {f_4^2- \frac{2}{3}f_{5t}} + \frac{\Delta^2}{12f_3^2},
\end{align}
\end{subequations}
where $f_4$ and $f_5$ are arbitrary functions of $t$.
The explicit solution for the mKP is given with substituting \eqref{slff} into \eqref{scon}
\begin{align}
& u=\frac{\Delta}{2f_3}\tan\Bigl( \frac{\Delta }{6f_3^2} (2f_{3t} y^2 + 6f_3f_4 y - 3f_3x - 3f_3f_4 )\Bigr) + \frac{2f_{3t} y}{3f_3} + f_4, \nonumber \\
& w= - \Delta(\frac{2}{3} \frac{f_{3t}}{f_3^2} y + \frac{f_4}{f_3})\tan\Bigl( \frac{\Delta }{6f_3^2} (2f_{3t} y^2 + 6f_3f_4 y - 3f_3x - 3f_3f_4 )\Bigr) + \frac{4f_{3t}^2 -4f_{3tt}f_3 }{9f_3^2} y^2 \\
& \hspace{0.7cm} - \frac{4}{3}f_{4t}y - \frac{2}{3} \frac{f_{3t}}{f_3}x - \frac{4f_{3t}f_5}{9f_3} + \frac{2}{3}f_{5t} - \frac{\Delta^2}{12f_3^2} + f_{4}^2.\nonumber
\end{align}

\section{CTE method for mKP system}

Recently, the consistent Riccati expansion (CRE)/consistent tanh expansion (CTE) methods are developed to find interaction solutions
between solitons and other types of nonlinear waves such as cnoidal waves, Painlev\'{e} waves and Airy waves \cite{gaox,cxp,cte,Interaction,wang}.
According to the CTE method, we assume the solution for the mKP equation \eqref{mkp}
has the generalized truncated tanh expansion form \cite{gaox,cxp}
\begin{align}\label{gen}
u = u_{0} + u_{1}\tanh (f), \hspace{1cm} w = w_{0} + w_{1}\tanh (f),
\end{align}
where $u_{0}$, $u_{1}$, $w_{0}$, $w_{1}$ and $f$ are functions of
$(x, y, t)$ and should be determined later.
By substituting \eqref{gen} into the mKP system \eqref{mkp} and vanishing coefficients of all the powers of $\tanh(f)$,
we can prove the following nonauto-BT theorem after some detail calculations.
\\
{\bf Nonauto-BT theorem.} If $f$ is the solution of the following equation
\begin{align}\label{fvet}
f_{xxxx} + 3 f_{yy} - 4f_{xt} - 4f_{xx}f_x^2 + \frac{4f_{t}f_{xx}} {f_{x}} - \frac{4f_{xx}f_{xxx}}{f_x} + \frac{3f_{xx}^3}{f_x^2} - \frac{3f_{xx}f_y^2}{f_x^2}=0,
\end{align}
then $u$ and $w$ with
\begin{align}\label{uvetor}
u  = - \frac{f_{xx}+f_y}{2f_x} + f_x\tanh(f), \hspace{0.3cm} w  = - \frac{1}{3}f_x^2 - \frac{2f_t}{3f_x} + \frac{f_{xxx}}{6f_x} - \frac{f_{xy}}{2f_x} - \frac{f_{xx}^2}{4f_x^2} + \frac{f_y^2}{4f_x^2} + f_y\tanh(f),
\end{align}
are a solution of the mKP system \eqref{mkp}.
Once the solutions of \eqref{fvet} are known, the corresponding
expression $u, w$ for \eqref{uvetor} can be obtained with the nonauto-BT theorem, whence the new solutions of \eqref{mkp} can be obtained.
Here we list three interesting examples.

A quite trivial solution of \eqref{fvet} has the form
\begin{align}\label{soll}
f = k x + ly + \omega t ,
\end{align}
where $k$, $l$ and $\omega$ are all the free constants.
Substituting the trivial solution \eqref{soll} into \eqref{uvetor}, the soliton solution of the mKP system yields
\begin{align}\label{fsolut}
& u = k \tanh(k x + ly + \omega t) - \frac{l}{2k},\\
& w = l \tanh(k x + ly + \omega t) - \frac{k^2}{3} -\frac{2\omega}{3k} + \frac{l^2}{4k^2}.
\end{align}

The soliton-cnoidal wave interaction solution for the mKP equation possesses the form
\begin{align}\label{sol2}
f = k x + ly + \omega t + F(k_1 x + l_1y + \omega_1 t),
\end{align}
where $k_1$, $l_1$ and $\omega_1$ are all the free arbitrary constants. Substituting \eqref{sol2} into \eqref{uvetor}, we have
\begin{align}\label{eqf}
F_{1X}^2 = 4F_1^4 + a_1 F_1^3 + a_2 F_1^2 + a_3 F_1 + a_4, \hspace{0.5cm} F_1=F_X, \hspace{0.5cm} F(k_1 x + l_1y + \omega_1 t) = F(X),
\end{align}
where
\begin{align}
& a_1= \frac{8k}{k_1} - 2c_2k_1^2, \hspace{1.2cm} a_3=-\frac{6kl_1^2}{k_1^5} + \frac{4k\omega_1 + 6l_1l}{k_1^4} - \frac{4\omega}{k^3} + 2k(2c_1-3c_2k), \nonumber \\
& a_2=(2c_1 - 6c_2k)k_1 + \frac{4k^2}{k_1^2},\hspace{0.5cm} a_4 = -\frac{5k^2 l_1^2}{k_1^6} + \frac{4k (k\omega_1+l_1l)}{k_1^5} +
\frac{l^2 - 4k\omega_2}{k_1^4} + \frac{2k^2 (c_1- c_2 k)}{k_1},\nonumber
\end{align}
and $c_1$ and $c_2$ are arbitrary constants. Similar the last section, the field $F$ has the form
\begin{align}\label{fief}
F = \frac{2}{\sqrt{(r_1-r_3)(r_2-r_4)}} \Bigl((r_3-r_4)E_{\pi} (S, \frac{r_1-r_4}{r_1-r_3}, m) - r_3E_F(S, m)\Bigr),\hspace{0.2cm} m=\sqrt{\frac{(r_1-r_4)(r_2-r_3)}{(r_1-r_3)(r_2-r_4)}},
\end{align}
where $S$ is the Jacobi elliptic function $S=\mathrm{sn}(\sqrt{(r_1-r_3)(r_2-r_4)}X, m)$, $E_F$ and $E_{\pi}$ are the first and third incomplete elliptic integrals and $r_1, r_2, r_3, r_4$ are related with $a_1, a_2, a_3, a_4$ in the following relations
\begin{align}
& a_1=-4(r_1+r_2+r_3+r_4), \hspace{0.9cm} a_2= 4(r_1r_2+r_1r_3+r_1r_4+r_2r_3+r_2r_4+r_3r_4), \nonumber  \\
& a_3 = -4(r_1r_2r_3+r_1r_2r_4+r_1r_3r_4+r_2r_3r_4), \hspace{0.9cm} a_4=4r_1r_2r_3r_4.\nonumber
\end{align}

The soliton-multiple wave interaction solution for the mKP equation has the following form
\begin{align}
f= k x + ly + \omega t + \frac{1}{2} \ln \bigl(1+\sum_{i=1}^n\exp(k_ix+l_iy+\omega_i t)\bigr),
\end{align}
where $k_i$ are arbitrary constants while $l_i$ and $\omega_i$ are determined
by the relations
\begin{align}\label{symb}
l_i = \frac{k_i}{2k} (4k^3+6k^2k_i + 2kk_i^2 \pm 6kl \pm 3k_il + 2\omega), \hspace{0.8cm} \omega_i=\pm \frac{k_i}{k}(2k^2+kk_i \pm l).
\end{align}
We can also find other relations to satisfy \eqref{fvet} and do not list it here. In the following, we select the symbol $``\pm"$ in \eqref{symb} as $``+"$ to plot the figures.
Figures 2 and 3 display the special interaction behavior between a soliton and one-resonant soliton solution with the parameters selected as $n=1, k=-1, k_1=1, l=0, \omega=2$ and $n=1, k=1, k_1=-1, l=1, \omega=2$ respectively. It demonstrate that the interaction behavior is different with selecting different parameters. Figure 4 displays the special interaction behavior between a soliton and two-resonant soliton solutions with the parameters selected as $n=2, k=1, k_1=1, k_2=-2, l=1, \omega=-1$.
\input epsf
\begin{figure}
\centering
\rotatebox{270}{\includegraphics[width=4.35cm]{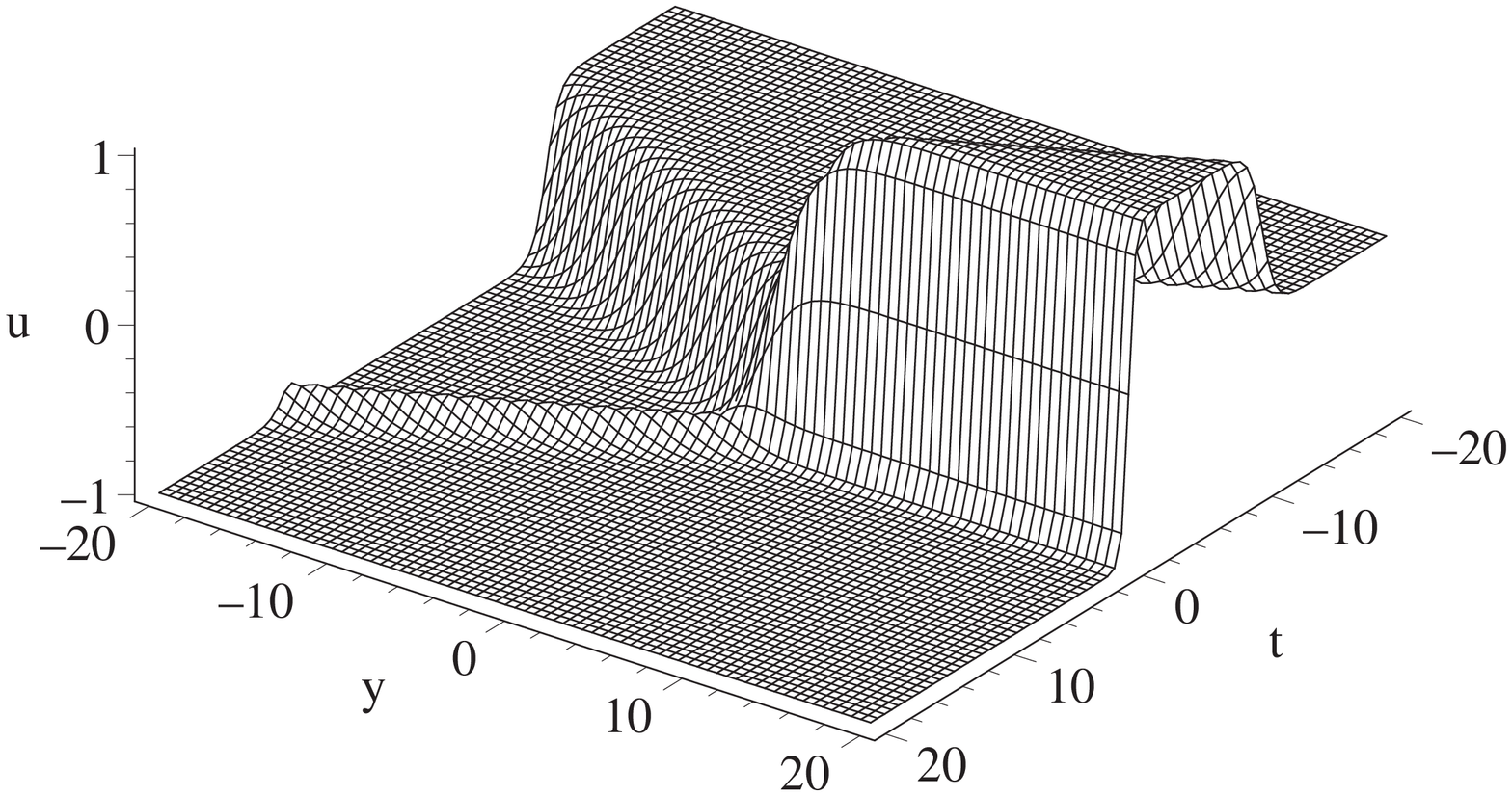}}
\rotatebox{270}{\includegraphics[width=4.85cm]{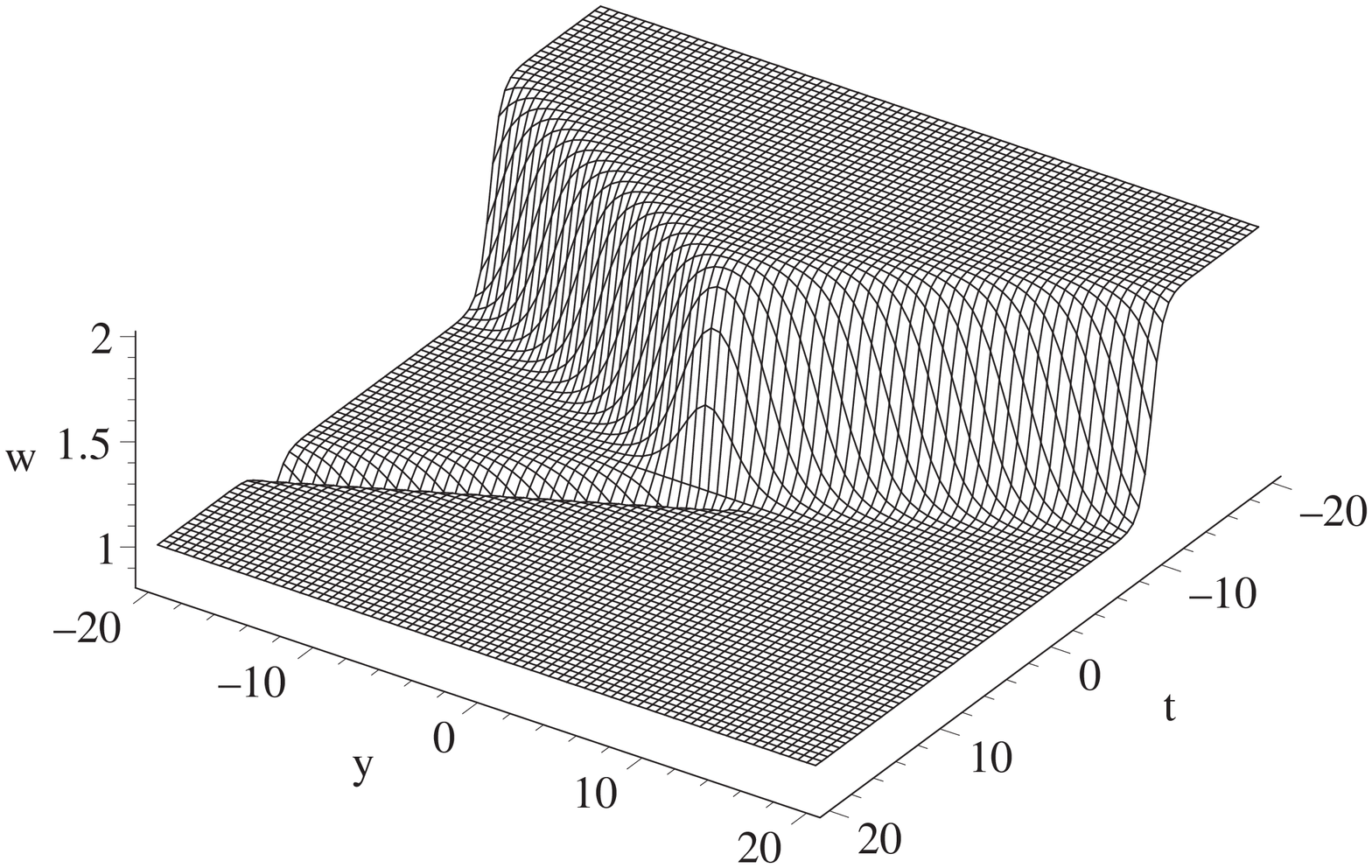}}
\caption{The interaction solution between a soliton and one-resonant soliton solution for fields $u$ and $w$ at $x=0$ respectively.
The parameters are $n=1, k=-1, k_1=1, l=0, \omega=2$.}
\end{figure}
\input epsf
\begin{figure}
\centering
\rotatebox{270}{\includegraphics[width=4.2cm]{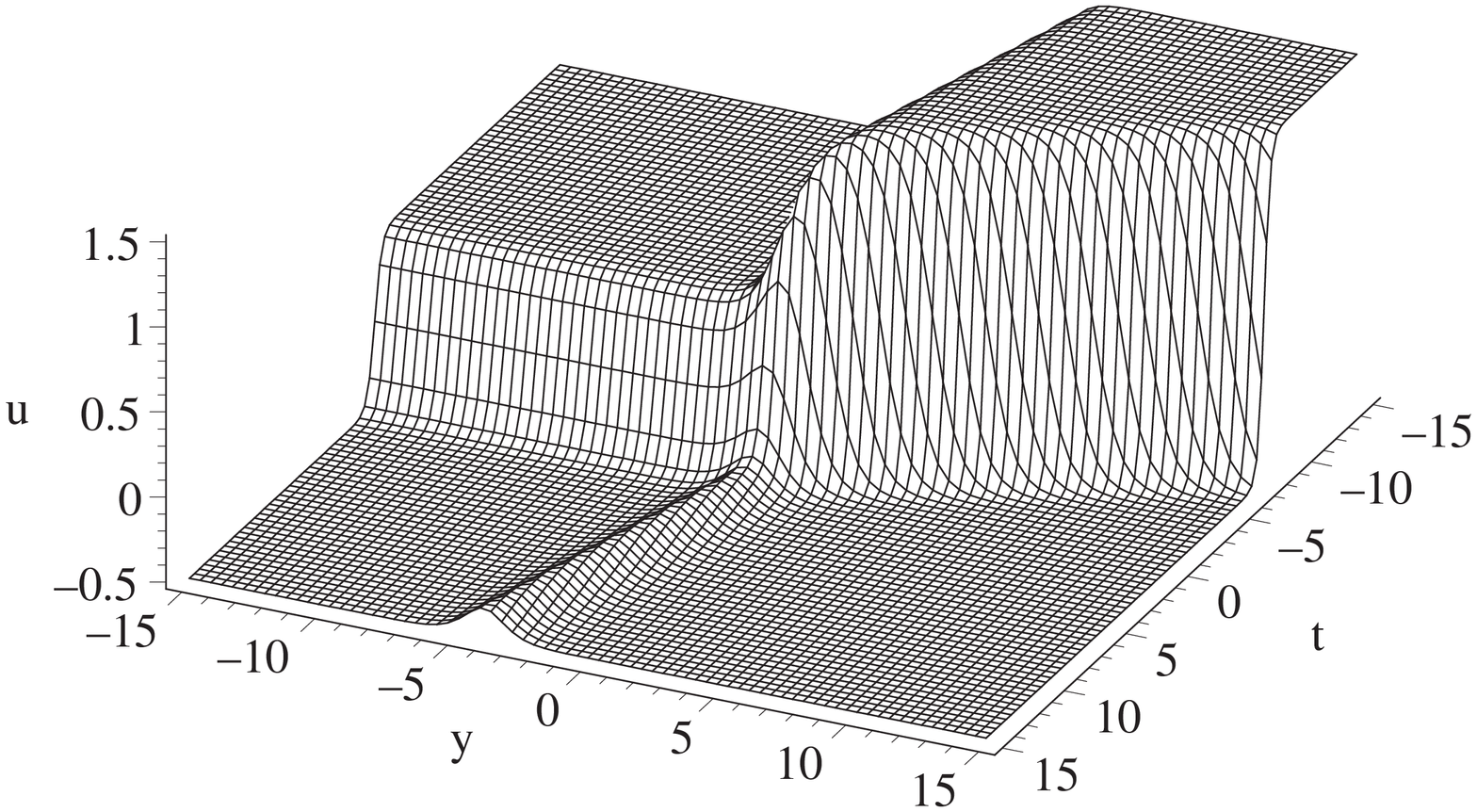}}
\rotatebox{270}{\includegraphics[width=4.75cm]{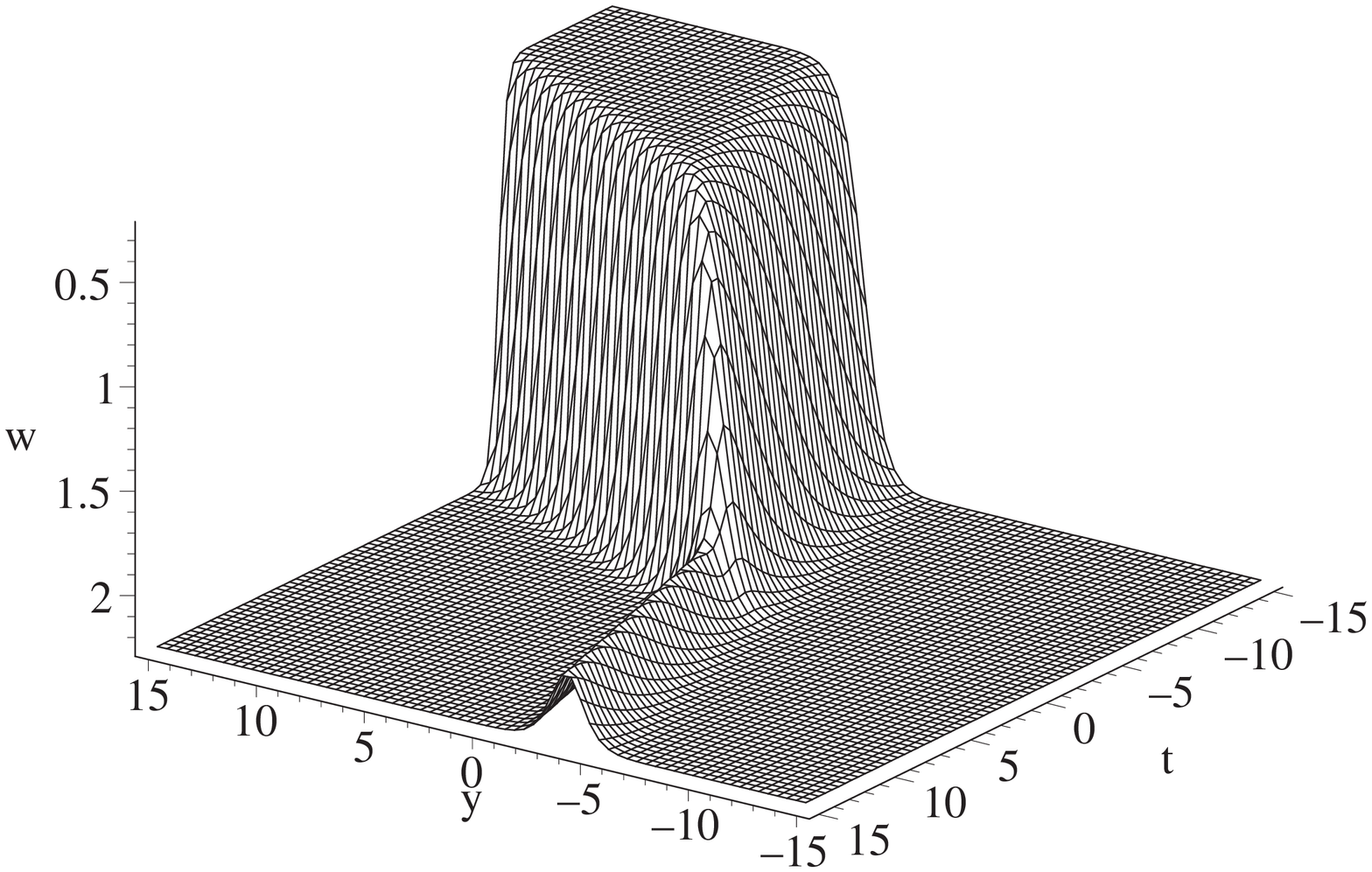}}
\caption{The interaction solution between a soliton and one-resonant soliton solution for fields $u$ and $w$ at $x=0$ respectively. The parameters are $n=1, k=1, k_1=-1, l=1, \omega=2$.}
\end{figure}
\input epsf
\begin{figure}
\centering
\rotatebox{270}{\includegraphics[width=3.65cm]{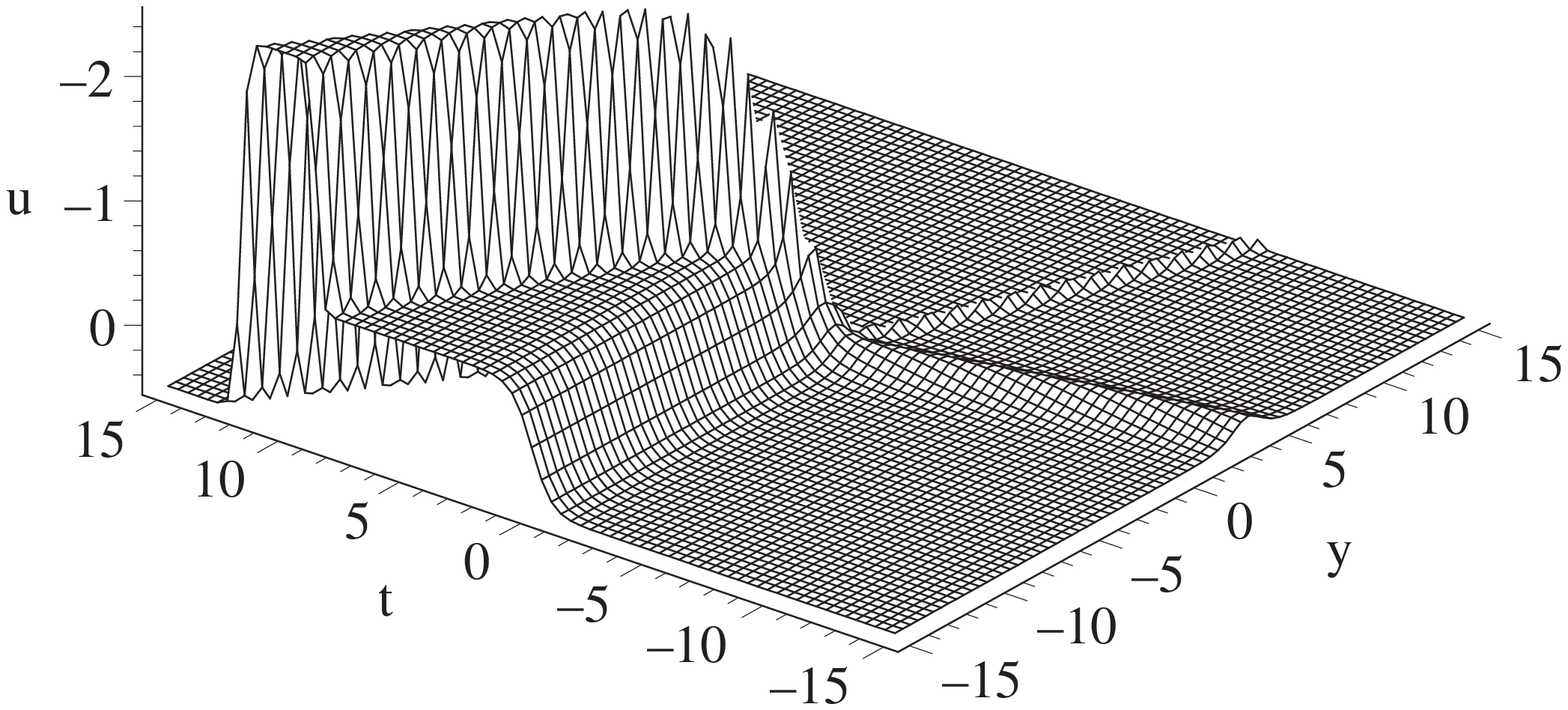}}
\rotatebox{270}{\includegraphics[width=3.8cm]{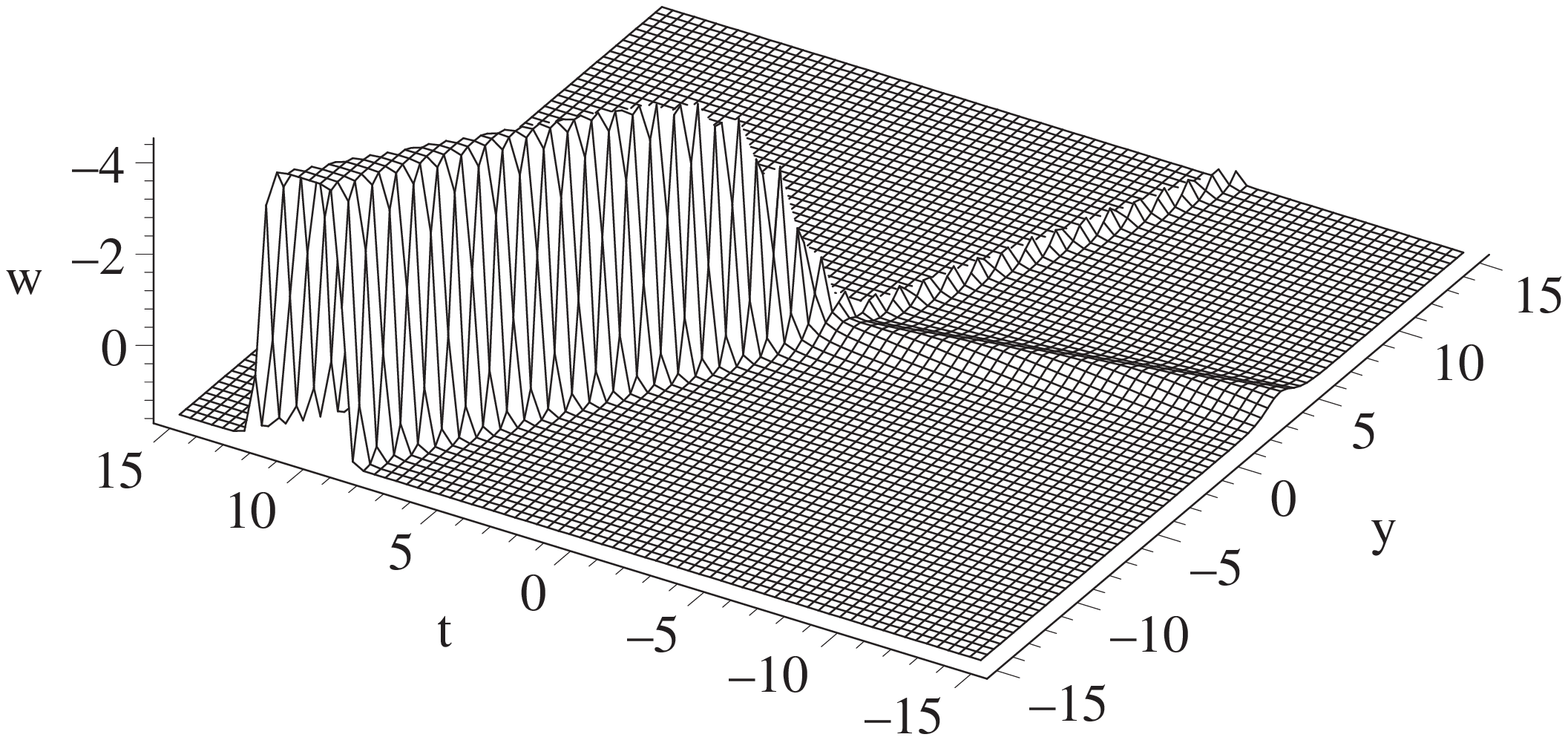}}
\caption{The interaction solution between a soliton and two-resonant soliton solutions for fields $u$ and $w$ at $x=0$ respectively. The parameters are $n=2, k=1, k_1=1, k_2=-2, l=1, \omega=-1$.}
\end{figure}

\section{Conclusions}

In summary, the nonlocal symmetries of the mKP equation are obtained with the truncated Painlev\'{e} method.
To solve the initial value problem related by the nonlocal symmetries, we prolong the mKP equation
such that nonlocal symmetries becomes the local Lie point symmetries for the prolonged system. The finite symmetry transformations of the prolonged
mKP system is derived by using the Lie's first principle. The multi-solitary wave solution for mKP equation is given with the finite symmetry transformations.
Thanks to the localization process, the nonlocal symmetries are used to find possible
symmetry reductions. The interaction solutions among one soliton and cnoidal waves are given as shown in \eqref{slfc} with \eqref{jorb}.
In the meanwhile, the CTE method is applied to the mKP equation.
With the help of the CTE method, we have found abundant interaction solutions among
a soliton and other types of nonlinear waves such as cnoidal periodic waves and multiple
resonant soliton solutions.
There exist other methods to construct the nonlocal symmetries such as those obtained from
the bilinear forms and negative hierarchies \cite{nonlo,nonloc},
nonlinearizations \cite{caoc}, point symmetries \cite{poin} and self-consistent sources \cite{zeng} etc. Using these various nonlocal symmetries to
derive new interaction solutions of the nonlinear integrable systems are worthy of further study.

{\bf Acknowledgment}:

\noindent
I would like to thank S. Y. Lou and J. Lin for useful discussions.
This work was partially supported by the National Natural Science Foundation of China under Grant (No. 11305106)
and the Natural Science Foundation of Zhejiang Province of China under Grant (No. LQ13A050001). The author is indebted
to the referees' useful comments.

\end{document}